\documentclass{article}
\usepackage{amssymb,amsmath}
\usepackage[]{algorithm2e}
\usepackage{graphicx}
\usepackage{authblk}
\usepackage{subcaption}
\usepackage[authoryear]{natbib}

\def\Ttheta{\boldsymbol\theta}

\def\bw{\mathbf w}
\def\by{\mathbf y}

\def\bs{\mathbf s}
\def\bc{\mathbf c}
\def\TX{\mathbf X}
\def\TQ{\mathbf Q}

\begin{document}
\title{Multi-resolution neural networks for \\tracking seismic horizons from few training images}

\author[1]{Bas Peters}
\author[2]{Justin Granek}
\author[1]{Eldad Haber}

\affil[1]{University of British Columbia, Vancouver, Canada}
\affil[2]{Computational Geosciences Inc.}
\date{November 29, 2018}
\maketitle

\begin{abstract}
Detecting a specific horizon in seismic images is a valuable tool for geological interpretation. Because hand-picking the locations of the horizon is a time-consuming process, automated computational methods were developed starting three decades ago. Older techniques for such picking include interpolation of control points however, in recent years neural networks have been used for this task. Until now, most networks trained on small patches from larger images. This limits the networks ability to learn from large-scale geologic structures. Moreover, currently available networks and training strategies require label patches that have full and continuous annotations, which are also time-consuming to generate. 

We propose a projected loss-function for training convolutional networks with a multi-resolution structure, including variants of the U-net. Our networks learn from a small number of large seismic images without creating patches. The projected loss-function enables training on labels with just a few annotated pixels and has no issue with the other unknown label pixels. Training uses all data without reserving some for validation. Only the labels are split into training/testing. Contrary to other work on horizon tracking, we train the network to perform non-linear regression, and not classification. As such, we propose labels as the convolution of a Gaussian kernel and the known horizon locations that indicate uncertainty in the labels. The network output is the probability of the horizon location. We demonstrate the proposed computational ingredients on two different datasets, for horizon extrapolation and interpolation. We show that the predictions of our methodology are accurate even in areas far from known horizon locations because our learning strategy exploits all data in large seismic images.
\end{abstract}

\section{Introduction}

\begin{figure}
 	\centering
 	\includegraphics[width=1.0\textwidth]{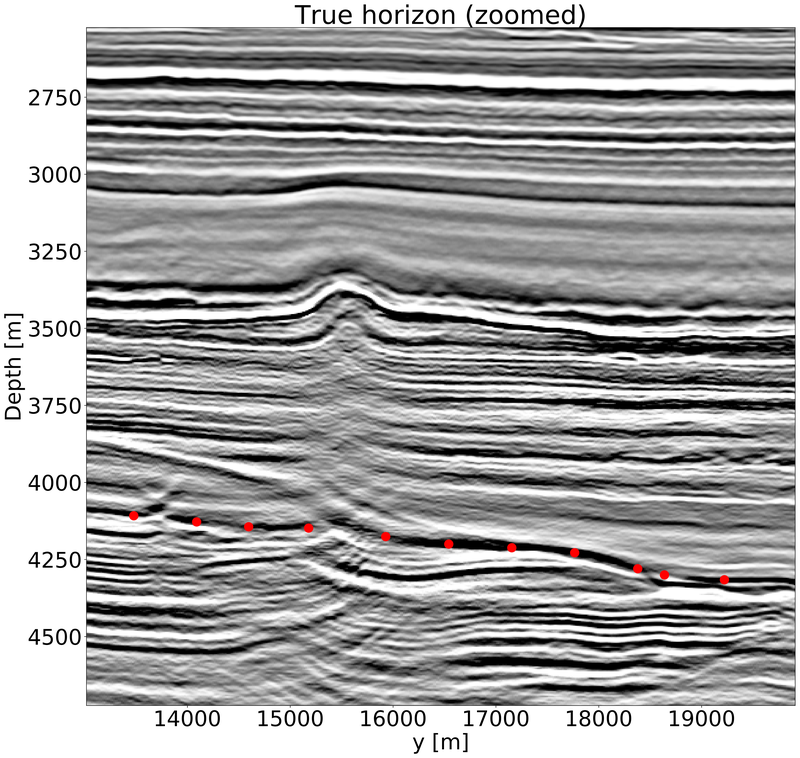}
 	\caption{A seismic image where red dots indicate the x-y-z locations of a horizon of interest. This image is a small part of the size we use for training.}
 	\label{fig:Figure1}
 \end{figure}

Geologic horizons are interfaces between two rock types with distinct petrophysical properties. These horizons are of great importance for understanding the geology and targeting resources such as hydrocarbons and water. Seismic imaging of the subsurface is the method of choice to obtain high-resolution images, from shallow to relatively large depths (see Figure \ref{fig:Figure1} for an example).

Seismic data is collected as a function of a shot and a recording location. The raw seismic data can be converted to images with the vertical axes representing either depth or time. In this work, we assume that the raw seismic data is already converted into such images. Traditionally, these seismic images are then manually interpreted by experts to identify and interpret the horizons. Hand picking the horizons in large seismic cross-sections or 3D volumes can be very time-consuming, especially where the resolution of the seismic image is sub-optimal or the geology is more complex. 

\subsection{Previous Work \& Related Problems}
Seismic attributes (e.g., pre-processed seismic images such as coherence, slope, energy, or dip) and automatic horizon tracking algorithms help seismic interpreters by tracking the horizon based on a few hand-picked reference locations, however, in more challenging environments can produce poor results and require added user supervision.

Automatic horizon detection algorithms come in various flavors for horizon interpretation. Neural network based methods have a long history for these applications. Early works \cite{A226265, Veezhinathan1993, doi:10.1190/1.1889749,doi:10.1190/1.1885963,  A1543200, doi:10.1190/1.1822449,doi:10.1190/1.1816150} use multi-layer perceptron or recurrent networks of a few layers. Neither the quantity or quality of data, nor the computing power used in these early works were comparable with today's standards. Some of the earlier references were limited to work with one or a few time-recordings (traces) at a time, thereby limiting the spatial information the networks can exploit. \cite{wu2018deep, doi:10.1190/segam2018-2997085.1} use a convolutional auto-encoder to perform segmentation of seismic images into a few regions. They pose the segmentation problem as a classification task where the horizons delineate the boundary between the class regions. Their training data is randomly selected out of a seismic volume and is, therefore, an example of interpolation of horizon locations. \cite{di2018developing} proposes to train on a large number of small annotated patches, for classifying seismic data volumes as an integrated geologic interpretation. A key difference from our work is that we work with the largest images practically possible, such that we can exploit spatial information over long distances to help predictions. A comparison by \cite{doi:10.1190/segam2018-2997085.1} confirms training image-to-image leads to better predictions compared to predicting the class of the central pixel from a small patch. 

There are also many algorithms for seismic horizon tracking that do not employ neural networks. These often require data pre-processing, or detect all horizons in a seismic image rather than one specific interface~\citep{doi:10.1190/1.1822449, A6184283}. \cite{doi:10.1190/geo2017-0830.1} propose a method that uses information over multiple length-scales on coarsened computational grids. 

Our goal of interpolating or extrapolating a specific seismic horizon is different from the related problems of salt-body \citep{doi:10.1190/tle37070529.1,doi:10.1190/segam2018-2997304.1}, fault \citep{doi:10.1111/j.1365-2478.2005.00489.x, doi:10.1190/tle36030208.1}, chimney detection \citep{doi:10.1190/1.1438976}, or multiple features \citep{doi:10.1190/segam2018-2997865.1}. For these applications, binary classification is the most common formulation: either pixels are the target of interest (i.e., salt) or they are not. Horizon detection presents a different challenge because every seismic image will contain many horizons, but we are typically interested in a small specific subset. We thus not only need to learn how to detect a horizon but also characteristics (thickness, amplitude, position in the stratigraphic sequence, depth, curvature) which help to uniquely identify it from other linear features in a seismic image.

\subsection{New contributions}

We provide a new approach to the horizon detection problem in seismic images. First, given the multiscale nature of seismic data, we employ a recently proposed network architecture \citep{Ronneberger2015} which has been shown to produce best-in-class performance for image segmentation in other fields such as medical imaging. Second, in contrast to the majority of datasets using deep learning for image recognition \citep{imagenetcvpr09,krizhevsky2009learning}, our dataset consists of a relatively small set of large images. To facilitate learning in such conditions, we introduce a partial loss function that enables training on partially labeled horizons. Our partial loss is different from methods that extract a small patch/cube around a label point (also known as a seed interpretation, \cite{doi:10.1190/1.1821262}) and classify the data patch by patch, yielding one classified pixel at a time. The partial loss enables us to train on sparse labels directly, without extracting a patch around the label point. 

Contrary to most work based on neural networks, we do not frame our problem as a classification task. Instead, we formulate non-linear regression problems where the label image values correspond to the probability of a horizon being at that depth for a given location. This is a convenient way to include uncertainty information on the horizon labels explicitly. The network output is therefore also an image that naturally conveys the uncertainty in the horizon depth estimates. Note that a classification approach provides the probability map of a class, which corresponds to the probability of a geologic rock type at each location in the image. The horizon location follows from such information as the points where the maximum class probability changes from one class to another, however, this does not directly provide the probability of the horizon at each location.

Because we train on large images, there is no need to create small patches. We thereby avoid manual user input on the window/patch size which would impact the results, as well as any artifacts resulting from a tiled solution to the problem. The dataset used consists of seismic images, and the algorithm is trained without the use of any other attribute images, wavelet information, or pre-processing that earlier work on horizon tracking used as supplemental input, see e.g. \cite{doi:10.1190/1.1438976, doi:10.1190/1.1484539, Leggett2003, A1543200, doi:10.1190/1.1816150}.

Finally, because our approach uses regression to train for the depth of the horizon, special consideration is necessary when preparing labels for the problem. We introduce a novel parametrization of the training label information which lends itself to a more transparent handling of uncertainty information and a probabilistic interpretation of the predicted results. Due to the sparsity of the horizon labels, the resulting training set can be very unbalanced. We handle this problem by re-balancing the training set at each iteration via per-class random sampling and demonstrate the importance of this step for the result.

\subsection{Application to field data}

We validate the proposed computational methods, loss function and learning strategy using seismic images from sedimentary areas in the North Sea and the Sea of Ireland. We demonstrate the effectiveness of our network architecture and new partial loss function, as well as investigate the difference between alternate problem setups, including interpolation versus extrapolation, and in-line versus cross-line predictions. 

\section{Label preparation and handling}

The raw label data are x-y-z coordinates of the location of the horizon of interest. We can directly plot the x-y-z coordinates in an image, by assigning the number $0$ to non-horizon locations and the number $1$ to horizon locations. We found training and predictions from this type of training labels rather ineffective, and it does not include valuable information about the uncertainty of the horizon picks.

The horizon picks are either hand-picked or obtained using an automatic horizon tracker with some human assistance and quality control. The selected x-y-z locations are therefore not completely accurate. Another source of label errors is the seismic image itself, from which the labels are generated. The quality of the seismic image decreases if there is noise in the data, or if the geology violates the assumptions on the migration method that generated the seismic image from raw seismic data. A common assumption is that the geology that is almost laterally invariant, i.e., slowly varying in the horizontal direction. More advanced imaging algorithms (e.g., reverse time-migration) assume a background velocity model that is approximately a smoothed version of the true velocity model. Violations of the assumptions result in parts of the seismic image becoming blurred, and continuous layers are broken up. The exact location of the horizon is ambiguous in these situations.

\begin{figure}
 	\centering
 	\begin{subfigure}[b]{1.0\textwidth}
 		\includegraphics[width=\textwidth]{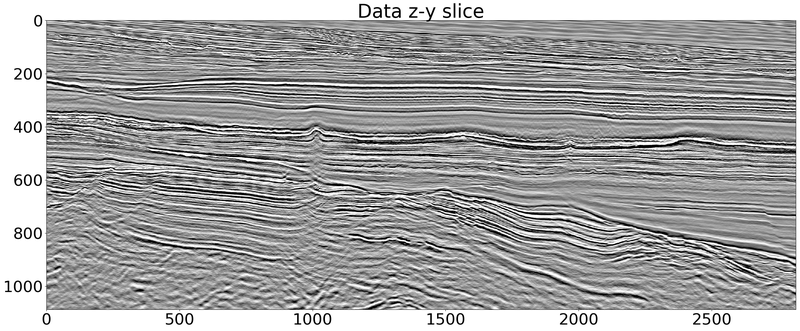}
 		\caption{}
 		\label{fig:Figure2a}
 	\end{subfigure}
 	\begin{subfigure}[b]{1.0\textwidth}
 		\includegraphics[width=\textwidth]{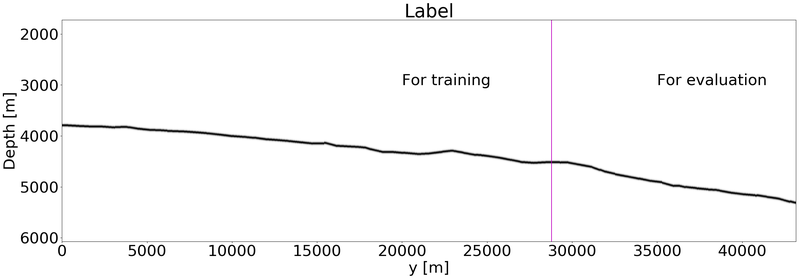}
 		\caption{}
 		\label{fig:Figure2b}
 	\end{subfigure}
 	\caption{a) A single data image. b) A single label image for horizon extrapolation. The part of the label image that is on the left of the vertical magenta line is used for training. The part that is right of the vertical line is for evaluation only. The label is the convolution of a Gaussian kernel with the horizon x-y-z locations.}
 \end{figure}

To reflect the uncertainty in the provided horizon labels, we add information about the uncertainty as follows: we convolve the horizon locations with a normalized Gaussian kernel. The resulting values are probabilities of the horizon location. The provided x-y-z location has the highest value, and the probability of a horizon tapers off as the distance from the x-y-z pick increases. In Figure \ref{fig:Figure2b} we show an example of a data image and label for the case we are given a horizon and need to extrapolate it. These images are of the size that we use for training.

\section{Network design}

A key component in the prediction of an interface is the network architecture. Most work done in the past uses very few layers for the prediction. It has been shown that for many vision applications such networks can have a limited power of prediction.
Recent architectures are based on deep networks that can contain tens if not hundreds of layers.
One such stable design is the residual network (ResNet, \cite{DBLPHeZRS15}) which can be written as
\begin{eqnarray}
\label{resnet}
{\bf y}_{j+1} = {\bf P}_j {\bf y}_{j} + g({\bf y}_j, {\boldsymbol \theta}_j).
\end{eqnarray}
Here, $\bf y$ are the layers in the network, $ {\boldsymbol \theta}$ are the parameters of each layer to be learned from the data and $g$ is a nonlinear function that consists of a hyperbolic tangent or a rectified linear unit (ReLu).
The transformation ${\bf P}$ is used to increase or decrease the number of channels of the network. For our problem, we start with a single channel (the seismic data) and open the network to a few tens of channels.

While ResNets have been very successful for image classification, they tend to be less accurate for segmentation problems. The main problem is scale; convolution is a local operator and therefore the network can have difficulties to learn features that span a number of scale-lengths.

In order to resolve this problem, we have used a U-Net \cite{Ronneberger2015} structure. U-Nets are similar to auto-encoders as they restrict (that is, downsample the image) as they go deeper. The network has two ``arms''. In the down-scale arm, equation \ref{resnet} is used, with a small modification
\begin{eqnarray}
\label{UnetDown}
{\bf y}_{j+1} = {\bf R}{\bf P}_j {\bf y}_{j} + {\bf R}g( {\bf y}_j, {\boldsymbol \theta}_j).
\end{eqnarray}
Here $\bf R$ is a restriction operator that down sample the image using a full weighting \citep{briggs}.

Let ${\bf y}_N$ be the image sampled on the lowest resolution.
In the second arm of the network the image is up-sampled to its original size, that is, the image is interpolated starting with ${\bf z}_N = {\bf y}_N$ by the equation:
\begin{eqnarray}
\label{UnetUp}
{\bf z}_{j-1} = {\bf y}_{j-1} + {\bf R}^{\top}{\bf P}_j {\bf z}_{j} + g({\bf R}^{\top} {\bf z}_j, {\boldsymbol \theta}_j^*)
\end{eqnarray}
Here, ${\bf R}^{\top}$ is the transpose of the restriction operator.
In order to obtain symmetry for the two branches of the net, we choose the parameters $\boldsymbol \theta^*$ of the up-going net to be the adjoints of the down going ones.
In particular, we use the transpose of the convolutions of the weights that are down-going.

The combination of low-resolution and high-resolution features allow the network to communicate between different scales, which is crucial for our application where reflectors have both local and global features.

\begin{table}[]
\centering
\begin{tabular}{r|r|c|c}
Layer \# &  Feature size &  \# of channels & kernel size  \\
\hline
1   & $1088 \times 2816$ & 4   & $3\times3$ \\
2   & $1088 \times 2816$ & 4   & $3\times3$ \\	
3   & $1088 \times 2816$ & 4   & $3\times3$ \\
4   & $544 \times 1408$   & 6   & $3\times3$ \\
5   & $544 \times 1408$   & 6   & $3\times3$ \\	
6   & $544 \times 1408$   & 6   & $3\times3$ \\
7   & $272 \times 704$     & 8   & $3\times3$ \\
8   & $272 \times 704$     & 8   & $3\times3$ \\
9   & $272 \times 704$     & 8   & $3\times3$ \\
10 & $136 \times 352$     & 12 & $3\times3$ \\
11 & $136 \times 352$     & 12 & $3\times3$ \\
12 & $136 \times 352$     & 12 & $3\times3$ \\
13 & $68 \times 176$	     & 16 & $3\times3$ \\
14 & $68 \times 176$	     & 16 & $3\times3$ \\
15 & $68 \times 176$	     & 16 & $3\times3$ \\
16 & $34 \times 88$	     & 24 & $3\times3$ \\
17 & $34 \times 88$	     & 24 & $3\times3$ \\
18 & $17 \times 44$	     & 32 & $3\times3$ \\
19 & $17 \times 44$	     & 32 & $3\times3$ \\
\end{tabular}
\caption{The down-scale arm of the U-net used for Sea of Ireland data. The up-sampled arm is the same in reverse order, and resuses the weights of the down-scale arm, resulting in $37336$ parameters. The network output has the same size as the input image, and represents the probability of the horizon of interest.}
\end{table}

\section{Partial loss function}

Consider a network $f(\by,\Ttheta) : \mathbb{R}^N \rightarrow \mathbb{R}^N$ that maps from (vectorized) images $\by \in \mathbb{R}^N$ of size $N = n_1 \times n_2$ to images of the same size. The network weights $\Ttheta$ are convolutional kernels, biases and a linear classifier $\bw \in \mathbb{R}^{n_\text{chan} \times 1}$ (in classification settings this would be a matrix). The last layer of the network reduces a tensor $\TX \in \mathbb{R}^{N \times n_\text{chan}}$ to $\bs \in \mathbb{R}^N$. The final network output is thus given by $\bs = \TX \bw$. We learn a single classifier $\bw$ that acts on every pixel of the image.

The least-squares loss is defined as
\begin{equation}\label{eq:LS}
	l(f(\by,\Ttheta),\bc) = \frac{1}{2}\| f(\by,\Ttheta) - \bc\|_2^2 ,
\end{equation}
where $\bc \in \mathbb{R}^{N}$ is a vectorized label image. This is a separable function, so we can compute a partial loss over a selection of pixels as
\begin{equation}\label{eq:LS_sub}
	l_\Omega(f(\by,\Ttheta),\bc) = \frac{1}{2} \sum_{i \in \Omega} (f(\by,\Ttheta)_i - \bc_i )^2,
\end{equation}
where $\Omega$ is the set of pixel indices where we have labels. Note that this is subsampling of the prediction, $f(\by,\Ttheta)$, which requires a full forward-pass through the network. The gradient computation uses the loss at the points in $\Omega$ only.

Another interpretation of the partial loss that is more common in geophysical literature is in terms of a projection. Define $\TQ \in \mathbb{R}^{n_\text{samp} \times N}$ as a projection matrix that projects onto the points in $\Omega$, i.e., $\TQ$ contains a subset of the rows of the identity matrix. We can then write the partial loss in equation \ref{eq:LS_sub} as
\begin{equation}\label{eq:PLS}
	l_{\TQ}(f(\by,\Ttheta),\bc) = \frac{1}{2}\| \TQ (f(\by,\Ttheta) - \bc_{\TQ}) \|_2^2.
\end{equation}
where $\bc_{\TQ}$ are the partial labels.
In this work we use the $\ell_1$ norm, which is separable as well. The partial, or projected $\ell_1$ loss is defined as
\begin{equation}\label{eq:L1_sub}
	l_\Omega(f(\by,\Ttheta),\bc) = \sum_{i \in \Omega} \left| f(\by,\Ttheta)_i - \bc_i \right|.
\end{equation}

The partial loss function enables us to train on partially known labels, as long as we know which pixels they are associated with, without labeling the whole seismic volume. 

\subsection{Stochastic optimization using a projected loss function}

Many neural network training strategies for classification of datasets that contain a large number of small ($\leq 128 \times 128$) images use random mini-batch stochastic gradient descent (SGD). At each iteration of SGD, the algorithm computes a gradient based on a small number of images and labels. For our applications, we typically only have access to a small number ($\leq 100$) of large images/labels ($\geq 1000 \times 1000$), sometimes even only a single image. If we were to compute a gradient based on a single image/label, there is only a single gradient and no stochastic effects. It has long been observed that full gradient methods are not competitive to randomized and stochastic gradient-based optimization algorithms for non-convex optimization in machine learning, particularly neural networks \citep{bottou2008tradeoffs}. The subsampling of the image and label pixels as proposed in the previous section provides us with a stochastic optimization algorithm by using a random subset of the points in $\Omega$ at each iteration.

\subsection{Re-balancing for the projected loss}

Seismic horizon detection problems have labels where most pixels have a value equal to $0$, which means there is no probability the horizon is located at that pixel. In each column, there are only a few entries that have a non-zero label value. This imbalance (about $30$ times more zero labels than non-zero in our numerical examples) can lead to slower training and low-quality predictions. To mitigate these issues, we apply binary re-balancing and use an equal number of zero and non-zero pixel values.

In a randomized stochastic optimization algorithm, at each iteration we draw $n_\text{samp}$ randomly selected samples out of the set of known label pixels $\Omega$. Binary re-balancing means there are $n_\text{samp}/2$ samples that have a label value equal to zero - denoted by the set $\mathcal{B}_1$ - and $n_\text{samp}/2$ samples that correspond to a non-zero label value - denoted by the set $\mathcal{B}_2$. The union of the two subsets is $\mathcal{A} = \mathcal{B}_1 \bigcup \mathcal{B}_2$.

We summarize the stochastic optimization algorithm for training neural networks using a partial loss function in combination with binary sample re-balancing in Algorithm \ref{alg:SGD_partial_eq}. The numerical examples show that balancing of zero and non-zero labels result in better predictions.\\

\begin{algorithm}[H]
	\textbf{Input}: \\
	$\by^1,\by^2,\cdots,\by^{n_\text{examples}}$    //data images \\
	$\bc^1,\bc^2,\cdots,\bc^{n_\text{examples}}$    //label images \\
	$\Omega^1,\Omega^2,\cdots,\Omega^{n_\text{examples}}$ //known label pixels in each image \\
	$n_\text{samp}$ //number of samples per iteration \\
	$\gamma$ //set learning rate \\
	\For{$1 : n_\text{epochs}$}{
	\For{$1 : n_\text{examples}$}{
	$\{\by,\bc,\Omega\}$  //draw random data, label image, known label indices \\
	$\mathcal{B}_1$  //draw at random $n_\text{samp}/2$ zero label indices from $\Omega$\\
	$\mathcal{B}_2$ //draw at random $n_\text{samp}/2$ non-zero label indices from $\Omega$\\
	$\mathcal{A} = \mathcal{B}_1 \bigcup \mathcal{B}_2$\\
	$l_\mathcal{A}(f(\by,\Ttheta),\bc)$, $\nabla_{\Ttheta} l_\mathcal{A}(f(\by,\Ttheta),\bc)$ //partial loss and gradient\\
	$\Ttheta \leftarrow \Ttheta - \gamma \nabla_{\Ttheta} l_\mathcal{A}(f(\by,\Ttheta),\bc)$ //update parameters
	}
	}
	\caption{Stochastic gradient descent based training algorithm with partial labels and binary random sample re-balancing.}
	\label{alg:SGD_partial_eq}
\end{algorithm}

\section{Field example of horizon tracking using neural networks}

Our data consists of seismic images that are models of the reflectivity of the Earth. The amplitude in the data relates to the elastic impedance contrast between the geological layers. The raw data have been processed into a large 3D model. We work with 2D slices. The labels are a combination of hand picking and algorithm assisted horizon tracking. Both data and picks were previously generated as part of a commercial exploration project by an external company.

We present results for extrapolation by in-line continuation, as well as interpolation from scattered horizon picks. The results also indicate the effect of balancing the number of zero and non-zero label values in each random batch.

We use the same network design for both examples and train the two networks using the projected $\ell_1$-loss as defined in equation \ref{eq:L1_sub}. The initialization of the network parameters is random.

\subsection{Horizon interpolation}

\begin{figure}
 	\centering
 	\includegraphics[width=1.0\textwidth]{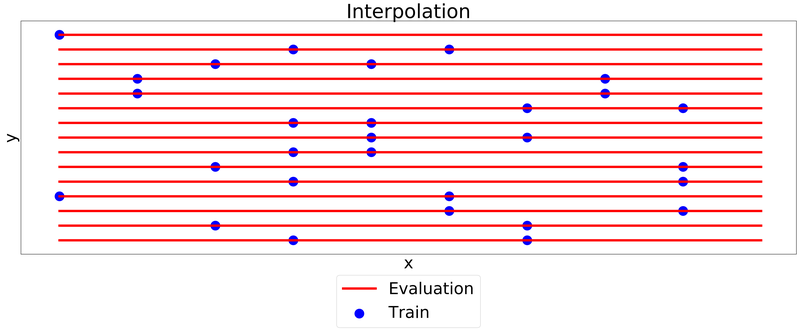}
 	\caption{Top view of horizon interpolation from scattered label points.}
 	\label{fig:Figure3}
 \end{figure}

\begin{figure}
     \centering
     \begin{subfigure}[b]{1.0\textwidth}
         \includegraphics[width=\textwidth]{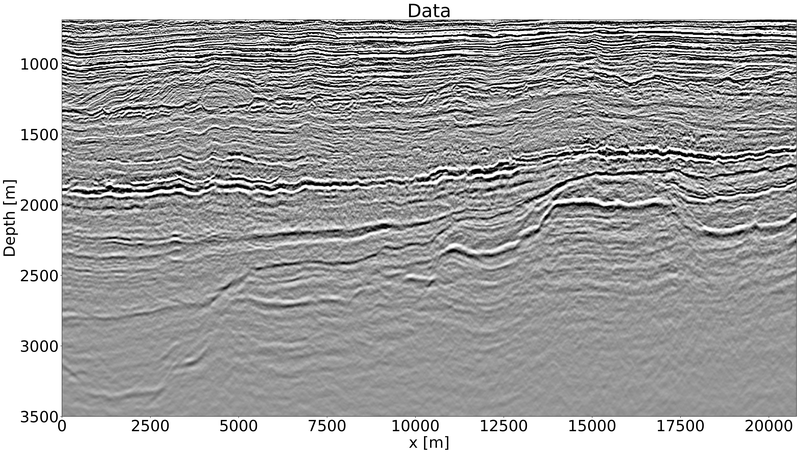}
         \caption{}
         \label{fig:Figure4a}
     \end{subfigure}
     \begin{subfigure}[b]{1.0\textwidth}
         \includegraphics[width=\textwidth]{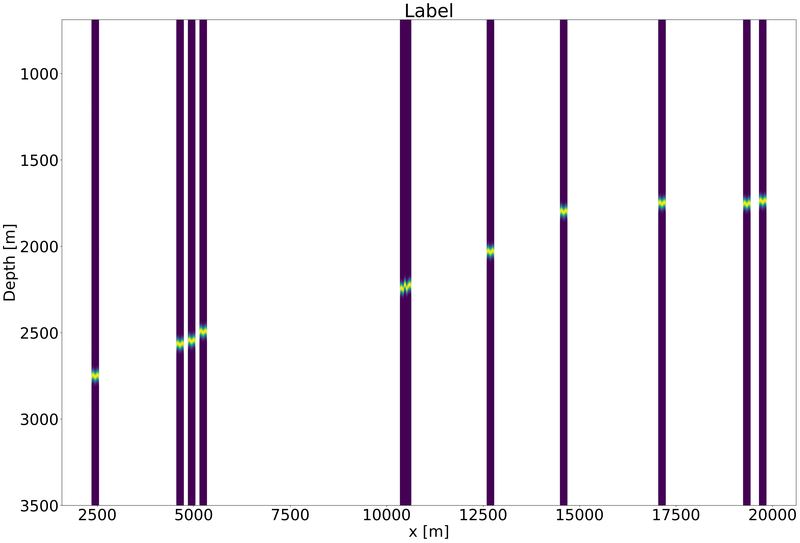}
         \caption{}
         \label{fig:Figure4b}
     \end{subfigure}
 	\caption{The data is the input for the network, the label is used in combination with the projected loss function that only computes the loss and gradient in the columns that have non-zero values. The non-zero label values are a Gaussian kernels centered at the provided horizon x-y-z locations -  white areas do not have associated label values.}
 \end{figure}

Hand-picking horizon locations is a time-consuming task. Many interpreted horizons have sparse spatial sampling as a result. In this case, we want to interpolate the picks to obtain continuous horizon surfaces, as shown in Figure \ref{fig:Figure3}. To be able to train on just a few labeled points in large images, we need a loss function that measures the loss at the labeled points only, but not at the other parts of the image. For a seismic horizon image, this means that we compute the loss based on the columns that have a horizon label (Figure \ref{fig:Figure4b}). In each of these columns, there is one horizon location; the other column entries are labels that indicate there is no horizon. The columns where we do not have any labeled information are excluded from training by the $\ell_1$ projected loss-function as defined in equation \ref{eq:L1_sub}; the network trains on all seismic data but only part of the label images.

The training data (Figure \ref{fig:Figure4a} ) are full 2D slices of size of $704 \times 1664$ pixels, without windowing or splitting into patches. The label images are only known at on average nine random locations per slice, provided by an industrial partner. We convolve the horizon location with a Gaussian kernel (in the vertical direction only) to assign an uncertainty to the hand-picked location. All other entries in the same column have a value equal to zero, which indicates the horizon does not occur at that location, see Figure \ref{fig:Figure4b}.

Training starts with $90$ epochs and a learning rate of $0.1$. Every iteration of each epoch uses a single data and label image. Out of the approximately $9 \times 704=6336$ known label pixels, we randomly select $150$ samples per iteration. As a result, not all label pixels are shown to the network during each epoch. We distribute the samples between zero and non-zero values equally. Note that the Gaussian kernel that we convolve with the horizon x-y-z locations has a width of $31$ pixels, so there are on average $9 \times 31$ non-zero label values per image. Training continues with another $65$ epochs and the learning rate is reduced by a factor of ten. The third and last training stage is $35$ epochs where we again reduce the learning rate by a factor of ten.

\begin{figure}
     \centering
     \begin{subfigure}[b]{1.0\textwidth}
         \includegraphics[width=\textwidth]{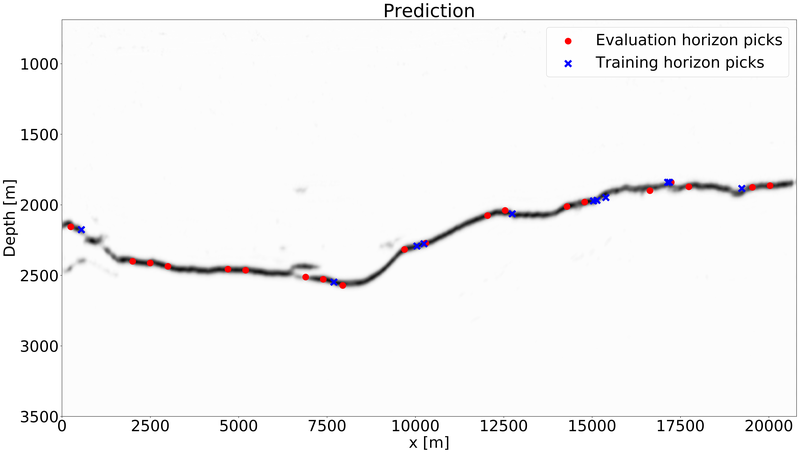}
         \caption{}
         \label{fig:Figure5a}
     \end{subfigure}
     \begin{subfigure}[b]{1.0\textwidth}
         \includegraphics[width=\textwidth]{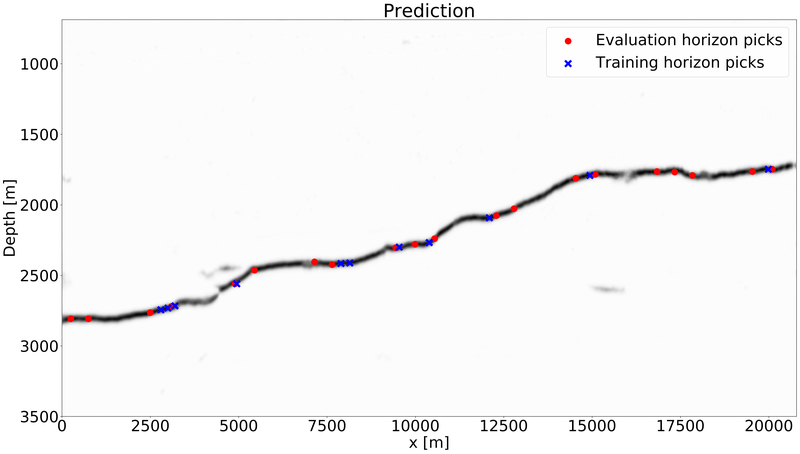}
         \caption{}
         \label{fig:Figure5b}
     \end{subfigure}
 	\caption{Predicted likelihood of the interface for two different slices.}
 \end{figure}

\begin{figure}
     \centering
     \begin{subfigure}[b]{1.0\textwidth}
         \includegraphics[width=\textwidth]{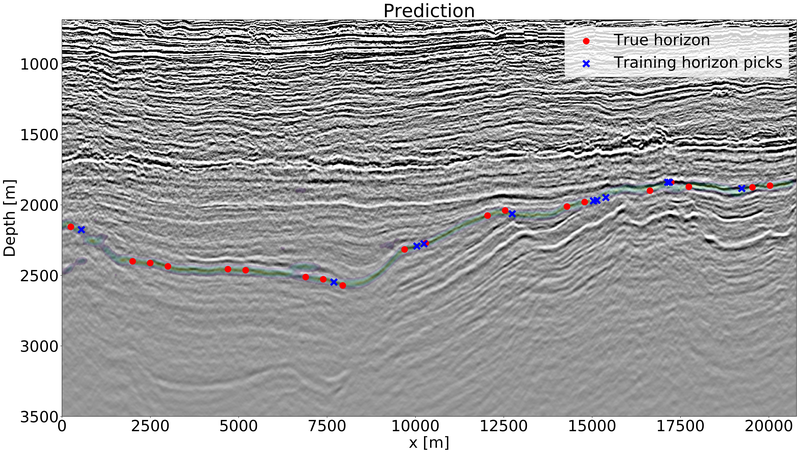}
         \caption{}
         \label{fig:Figure6a}
     \end{subfigure}
     \begin{subfigure}[b]{1.0\textwidth}
         \includegraphics[width=\textwidth]{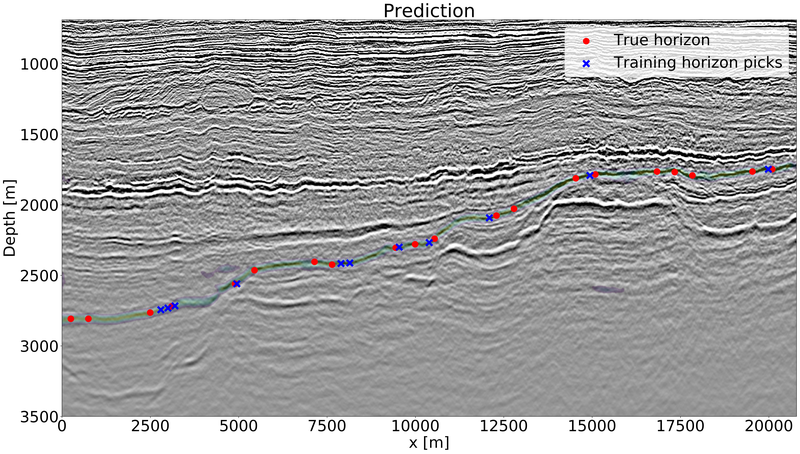}
         \caption{}
         \label{fig:Figure6b}
     \end{subfigure}
 	\caption{Predicted likelihood of the interface for two different slices overlaid on the seismic data.}
 \end{figure}

\begin{figure}
 	\centering
 	\includegraphics[width=1.0\textwidth]{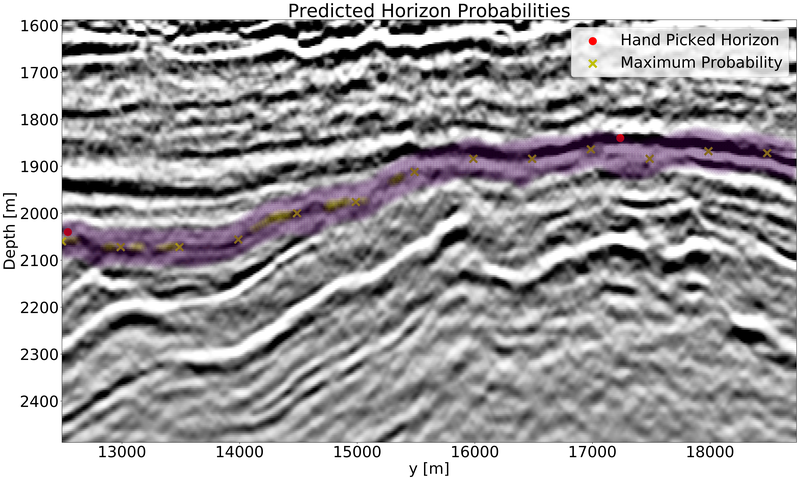}
 	\caption{Zoomed version of Figure \ref{fig:Figure6b}.}
 	\label{fig:Figure7}
 \end{figure}

Figure \ref{fig:Figure5b} displays the prediction for two slices. Figure \ref{fig:Figure6b} shows the same information using color-coding for the predicted probability and overlaid on the data. The zoomed version in Figure \ref{fig:Figure7} shows more details. 

\begin{figure}
 	\centering
 	\includegraphics[width=1.0\textwidth]{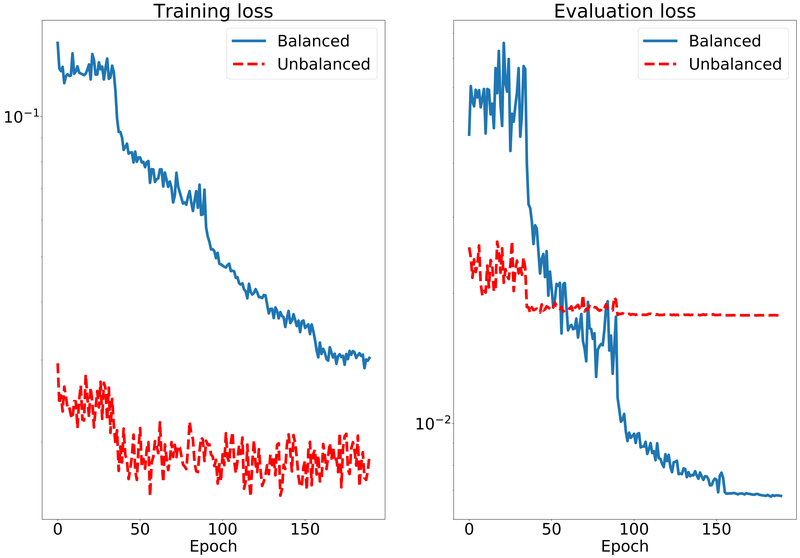}
 	\caption{Training and evaluation loss for training with and without re-balancing the number of zero and non-zero label samples. The training loss is much smaller without re-balancing because most of the label samples have a value equal to zero. The evaluation loss is measured over all available labels and it is evident that re-balancing leads to better evaluation performance.}
 	\label{fig:Figure8}
 \end{figure}

The results in Figures \ref{fig:Figure5b}, \ref{fig:Figure6b}, and \ref{fig:Figure7} show excellent predictions. The network output displays the probability of a horizon directly and no additional post-processing was applied. The average of nine picks per slice is not a lower limit or recommended number. Getting good predictions using fewer picks is possible. We point out that we could train more to reduce the validation loss, see Figure \ref{fig:Figure8}. We also did not use any data-augmentation, which could benefit the training in the case of fewer label points.

\begin{figure}
 	\centering
 	\includegraphics[width=1.0\textwidth]{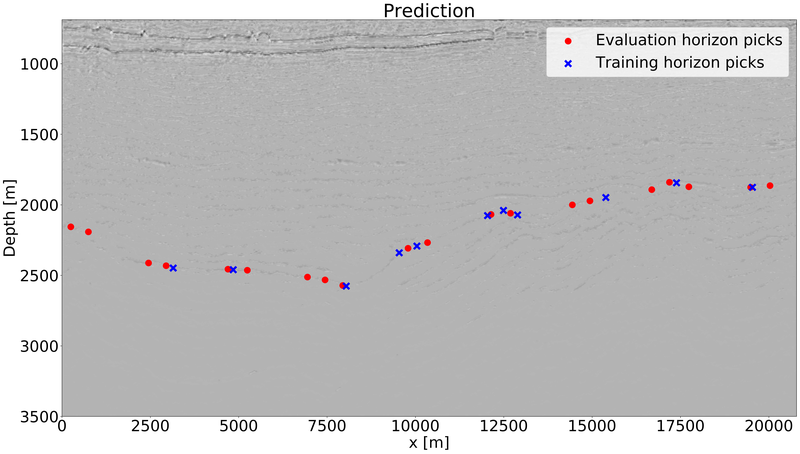}
 	\caption{Prediction when the network is trained without re-balancing the number of zero and non-zero label values.}
 	\label{fig:Figure9}
 \end{figure}

With regards to the balancing procedure outlined in an earlier section, the loss function logs in Figure \ref{fig:Figure8} clearly show that not balancing the number of zero and non-zero label points during each SGD iteration leads to a worse validation loss. Note that contrary to many works on horizon tracking using neural networks based on classification, our non-linear regression strategy does not have a prediction accuracy. Figure \ref{fig:Figure9} shows a prediction from training without balancing, which is not close to the desired output in any way.

\subsection{Horizon extrapolation}
Points that indicate the x-y-z locations of a horizon are also called horizon picks. Given a collection of picks in an area, we can try to extrapolate the horizon away from the known locations. Much historical industrial work produced large quantities of horizon picks that we can use for training. A potential challenge is that the extrapolation can be in areas with different geology than where the training picks are.

\begin{figure}
 	\centering
 	\includegraphics[width=1.0\textwidth]{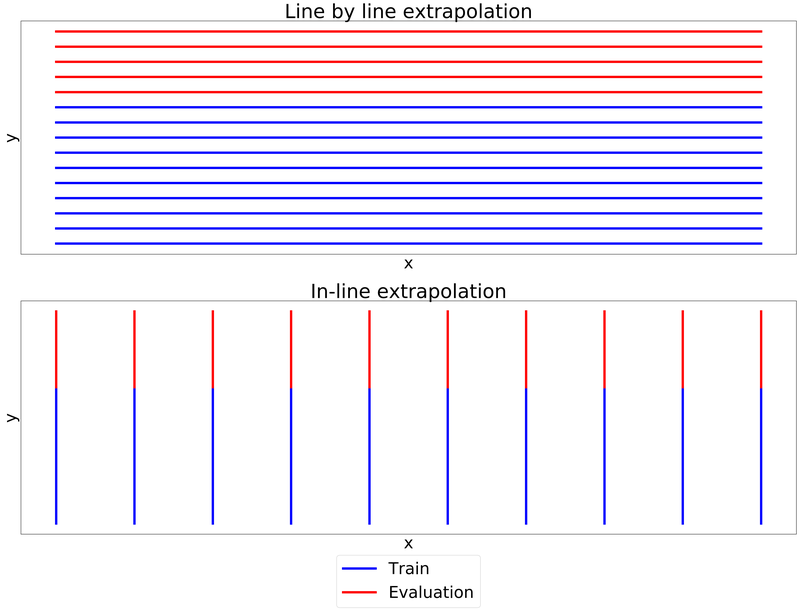}
 	\caption{Plan view of two different types of horizon extrapolation.}
 	\label{fig:Figure10}
 \end{figure}

There are multiple types of extrapolation, two of which are shown in Figure \ref{fig:Figure10}. Perhaps most similar to standard classifications or segmentation tasks on data sets containing many small images (e.g., MNIST, CIFAR), is to train on one set of images, then apply the trained network and classifier on another test set of images. We call this line-by-line or slice-by-slice learning. A slice refers to a 2D slice from a 3D tensor. The second strategy extrapolates a horizon in-line. The training procedure sees the full data (seismic image), but the label is only partially known.

\subsubsection{Line-by-line versus in-line extrapolation}
We provide some insight about which of the two types of horizon extrapolation is preferable. First of all, it is important to realize that the interpretation of seismic images is different from other problems, such as segmentation of images from video for self-driving vehicles. That application has pre-recorded video/images along with segmented labels available for training. The testing data arrives in real-time and segmentation needs to happen in a short amount of time. In our case, the complete seismic 3D volume is available at the time of training. It is only the labels that are incomplete. Therefore, we would like to use all training data, together with the labels corresponding to a part of the training data. In-line extrapolation keeps a number of slices separately for testing, so the network never has access to those seismic images. Contrary, in-line extrapolation trains on the full seismic slices, but sees only part of the labels, see Figure \ref{fig:Figure2b}. Because we will use a deep neural network with multiple convolutional layers and subsampling/upsampling stages, the data from the area without labels will influence the prediction in the area where we do have labels. For this reason, in-line extrapolation has the capability to utilize all data, and we focus on this method in the remainder of this paper.

For training, we use just $24$ images of size $1088 \times 2816$ pixels. There are three training stages. We start with $40$ epochs and a learning rate of $0.1$, followed by $30$ epochs with the learning rate reduced by a factor of ten. The last stage is another $20$ epochs where we reduce the learning rate by another factor of ten.

\begin{figure}
 	\centering
 	\begin{subfigure}[b]{1.0\textwidth}
 		\includegraphics[width=\textwidth]{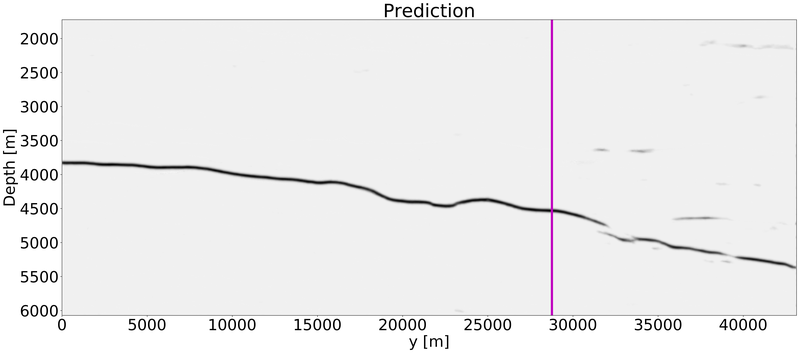}
 		\caption{}
 		\label{fig:Figure11a}
 	\end{subfigure}
 	\begin{subfigure}[b]{1.0\textwidth}
 		\includegraphics[width=\textwidth]{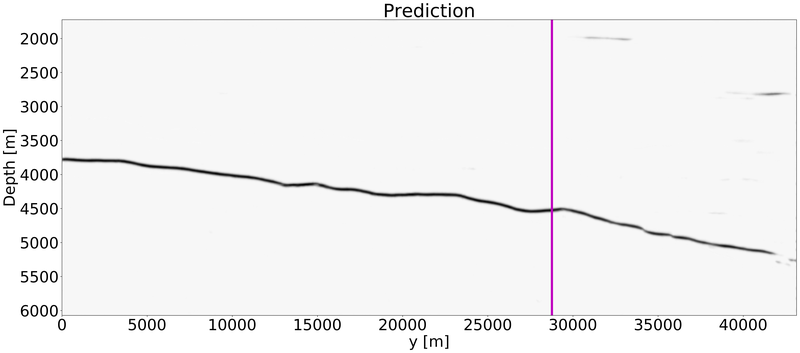}
 		\caption{}
 		\label{fig:Figure11b}
 	\end{subfigure}
 	\caption{Predictions for two slices on both training and evaluation data. The part of the figures left of the vertical line is the prediction on training data, on the right is the prediction on evaluation data.}
 \end{figure}

\begin{figure}
 	\centering
 	\begin{subfigure}[b]{1.0\textwidth}
 		\includegraphics[width=\textwidth]{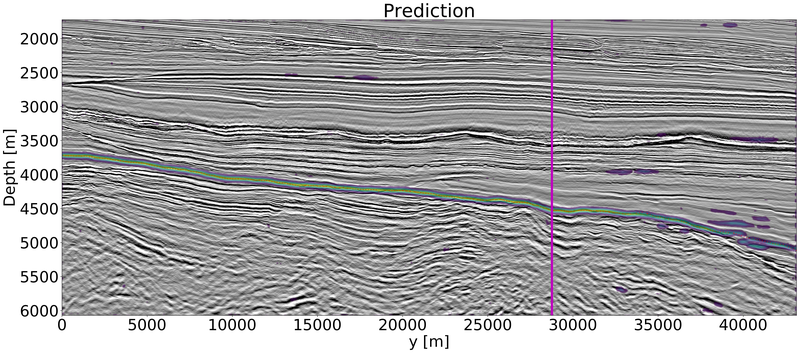}
 		\caption{}
 		\label{fig:Figure12a}
 	\end{subfigure}
 	\begin{subfigure}[b]{1.0\textwidth}
 		\includegraphics[width=\textwidth]{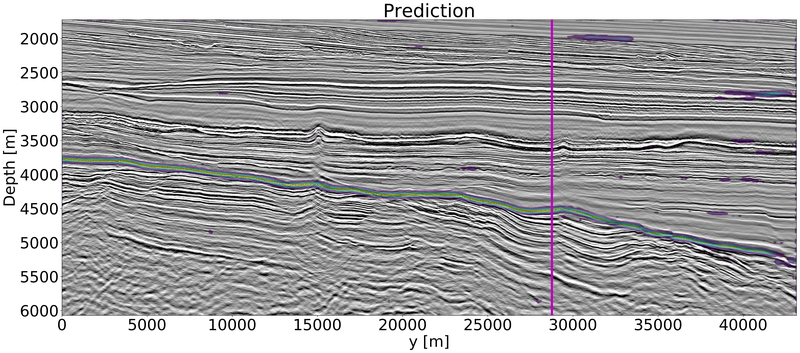}
 		\caption{}
 		\label{fig:Figure12b}
 	\end{subfigure}
 	\caption{Color-coded predictions overlaid on the seismic data for two slices of both training and evaluation data. Yellow colors indicate a high probability of an interface, purple corresponds to low probability. The true horizon location is indicated by a red dotted line. The part of the figures left of the vertical line is the prediction on training data, on the right is the prediction on evaluation data.}
 \end{figure}

\begin{figure}
 	\centering
 	\begin{subfigure}[b]{0.4\textwidth}
 		\includegraphics[width=\textwidth]{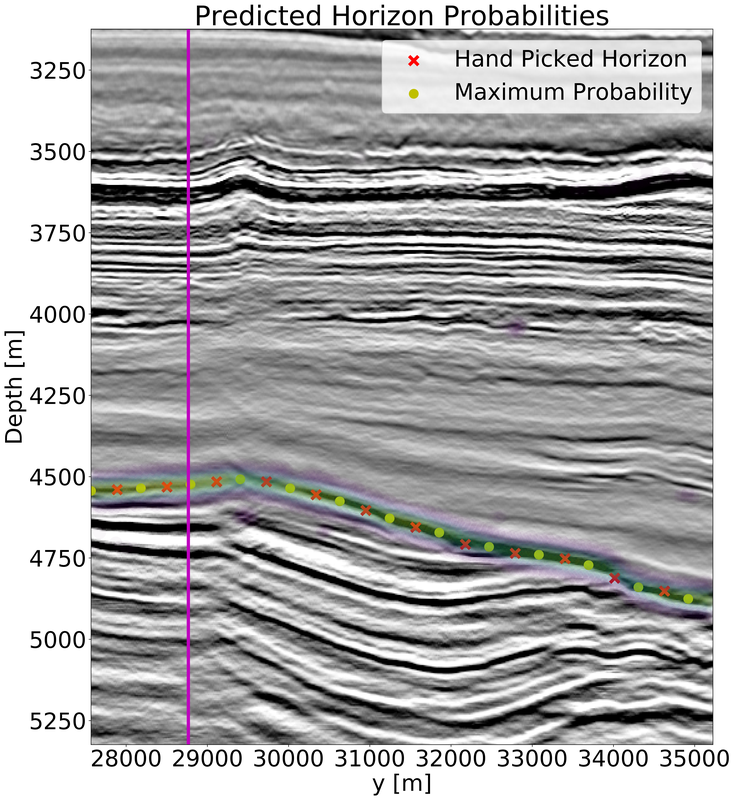}
 		\caption{}
 		\label{fig:Figure13a}
 	\end{subfigure}
 	\begin{subfigure}[b]{0.4\textwidth}
 		\includegraphics[width=\textwidth]{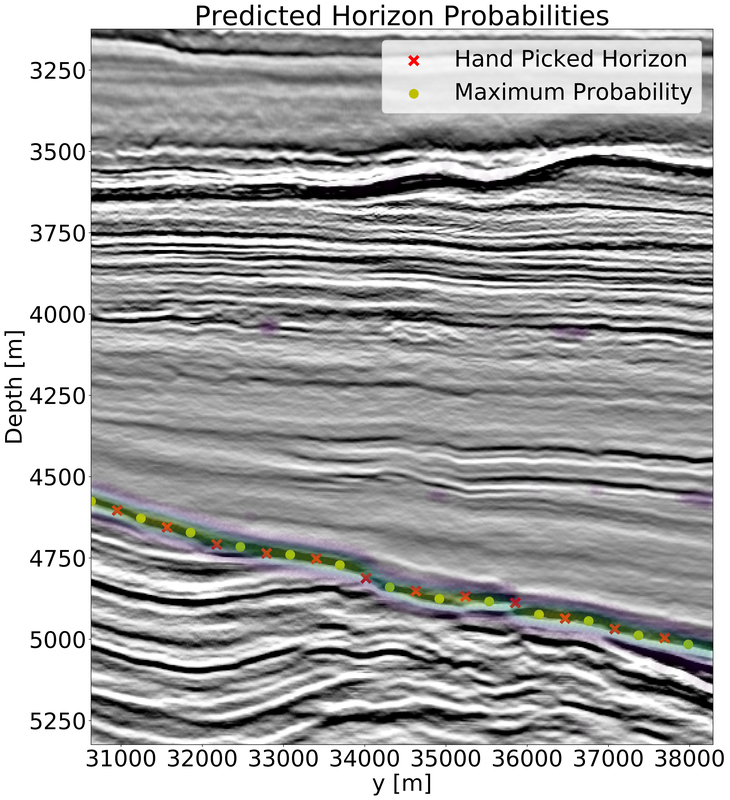}
 		\caption{}
 		\label{fig:Figure13b}
 	\end{subfigure}
 	\caption{Zoomed section of Figure \ref{fig:Figure12b}. Figure shows predicted probabilities, maximum predicted probability and the horizon locations that are hand-picked by an industrial partner.}
 \end{figure}

Figure \ref{fig:Figure2b} shows an example of the labels and data. The evaluation part of each data image is about $1/3$, which is the extrapolation distance of interest to an industrial partner. In Figure \ref{fig:Figure11b} we display the predictions for two slices. The prediction on the right of the vertical line shows that we generally predict a continuous line, but it is difficult to see how accurate the prediction is. By color coding the predicted likelihood of the horizon in Figure \ref{fig:Figure12b}, we see that the incorrectly highlighted areas have a much lower probability than the correct horizon locations. We also observe that our prediction on the training part is almost perfect. Figure \ref{fig:Figure13b} shows a zoomed-in section that better illustrates the relation between predicted probability and the seismic image.

\section{Conclusions}

In this work we provided a new look at the problem of detecting horizons in seismic images using neural networks. Specifically, we addressed extrapolation away from previously interpreted horizons, as well as the interpolation of a small number of scattered hand-picked horizon locations. The proposed networks, loss function, and learning strategies to overcome issues that limit the success of automatic interpretation using neural networks. We employ deep networks with a multi-resolution structure to train on a small number of large seismic images that take large-scale geological structures into account, in the sense that information propagates over long distances on multiple scales. This is not directly possible using standard network-based learning methods that train on small image patches. We proposed a projected loss function that enables training on label images with only a few annotated pixels. Generating such labels is easier and faster than working with conventional label images that need complete labeling of a full image or patch. The standard practice of splitting data and labels into training and test sets is no longer necessary when we train with the projected loss. In seismic imaging, we have access to all data during training. It is the labels that are incomplete. Our networks train on all available seismic images, and we compute the loss and gradient based on a small number of known label pixels. The data in areas without corresponding labels is still seen by the network, and because the network has multiple layers on multiple resolutions, the information influences the predictions and misfit at locations where we do have horizon picks. Application of the proposed network, loss function, and learning strategy to horizon extrapolation and interpolation showed that our methods provide accurate predictions and uncertainty estimates both close and farther from known horizon locations. Our experiments so far were restricted to sedimentary geological settings in the North Sea and Sea or Ireland. The proposed methods make automatic horizon detection possible using fewer horizon picks and take all available seismic data into account.

\newpage
\bibliographystyle{abbrvnat}
\bibliography{draft_V1_plain}{}

\begin{thebibliography}{27}
\providecommand{\natexlab}[1]{#1}
\providecommand{\url}[1]{\texttt{#1}}
\expandafter\ifx\csname urlstyle\endcsname\relax
  \providecommand{\doi}[1]{doi: #1}\else
  \providecommand{\doi}{doi: \begingroup \urlstyle{rm}\Url}\fi

\bibitem[Alaudah et~al.(2018)Alaudah, Gao, and
  AlRegib]{doi:10.1190/segam2018-2997865.1}
Y.~Alaudah, S.~Gao, and G.~AlRegib.
\newblock Learning to label seismic structures with deconvolution networks and
  weak labels.
\newblock In \emph{SEG Technical Program Expanded Abstracts 2018}, pages
  2121--2125, 2018.
\newblock \doi{10.1190/segam2018-2997865.1}.
\newblock URL
  \url{https://library.seg.org/doi/abs/10.1190/segam2018-2997865.1}.

\bibitem[Alberts et~al.(2000)Alberts, Warner, and
  Lister]{doi:10.1190/1.1816150}
P.~Alberts, M.~Warner, and D.~Lister.
\newblock Artificial neural networks for simultaneous multi horizon tracking
  across discontinuities.
\newblock In \emph{SEG Technical Program Expanded Abstracts 2000}, pages
  651--653. Society of Exploration Geophysicists, 2000.

\bibitem[Araya-Polo et~al.(2017)Araya-Polo, Dahlke, Frogner, Zhang, Poggio, and
  Hohl]{doi:10.1190/tle36030208.1}
M.~Araya-Polo, T.~Dahlke, C.~Frogner, C.~Zhang, T.~Poggio, and D.~Hohl.
\newblock Automated fault detection without seismic processing.
\newblock \emph{The Leading Edge}, 36\penalty0 (3):\penalty0 208--214, 2017.
\newblock \doi{10.1190/tle36030208.1}.
\newblock URL \url{https://doi.org/10.1190/tle36030208.1}.

\bibitem[Bottou and Bousquet(2008)]{bottou2008tradeoffs}
L.~Bottou and O.~Bousquet.
\newblock The tradeoffs of large scale learning.
\newblock In \emph{Advances in neural information processing systems}, pages
  161--168, 2008.

\bibitem[Briggs et~al.(2000)Briggs, Henson, and McCormick]{briggs}
W.~Briggs, V.~Henson, and S.~McCormick.
\newblock \emph{A Multigrid Tutorial, Second Edition}.
\newblock Society for Industrial and Applied Mathematics, second edition, 2000.
\newblock \doi{10.1137/1.9780898719505}.
\newblock URL \url{https://epubs.siam.org/doi/abs/10.1137/1.9780898719505}.

\bibitem[Deng et~al.(2009)Deng, Dong, Socher, Li, Li, and
  Fei-Fei]{imagenetcvpr09}
J.~Deng, W.~Dong, R.~Socher, L.~Li, K.~Li, and L.~Fei-Fei.
\newblock Imagenet: A large-scale hierarchical image database.
\newblock In \emph{2009 IEEE Conference on Computer Vision and Pattern
  Recognition}, pages 248--255, June 2009.
\newblock \doi{10.1109/CVPR.2009.5206848}.

\bibitem[Di(2018)]{di2018developing}
H.~Di.
\newblock Developing a seismic pattern interpretation network (spinet) for
  automated seismic interpretation.
\newblock \emph{arXiv preprint arXiv:1810.08517}, 2018.

\bibitem[Harrigan et~al.(1992)Harrigan, Kroh, Sandham, and Durrani]{A226265}
E.~Harrigan, J.~R. Kroh, W.~A. Sandham, and T.~S. Durrani.
\newblock Seismic horizon picking using an artificial neural network.
\newblock In \emph{[Proceedings] ICASSP-92: 1992 IEEE International Conference
  on Acoustics, Speech, and Signal Processing}, volume~3, pages 105--108 vol.3,
  March 1992.
\newblock \doi{10.1109/ICASSP.1992.226265}.

\bibitem[He et~al.(2015)He, Zhang, Ren, and Sun]{DBLPHeZRS15}
K.~He, X.~Zhang, S.~Ren, and J.~Sun.
\newblock Deep residual learning for image recognition.
\newblock \emph{CoRR}, abs/1512.03385, 2015.
\newblock URL \url{http://arxiv.org/abs/1512.03385}.

\bibitem[Huang(2005)]{doi:10.1190/1.1885963}
K.-Y. Huang.
\newblock Hopfield neural network for seismic horizon picking.
\newblock In \emph{SEG Technical Program Expanded Abstracts 1997}, pages
  562--565, 2005.
\newblock \doi{10.1190/1.1885963}.
\newblock URL \url{https://library.seg.org/doi/abs/10.1190/1.1885963}.

\bibitem[Huang et~al.(2005)Huang, Chang, Hsieh, Hsieh, Wang, and
  Tsai]{A1543200}
K.-Y. Huang, C.-H. Chang, W.-S. Hsieh, S.-C. Hsieh, L.~K. Wang, and F.-J. Tsai.
\newblock Cellular neural network for seismic horizon picking.
\newblock In \emph{2005 9th International Workshop on Cellular Neural Networks
  and Their Applications}, pages 219--222, May 2005.
\newblock \doi{10.1109/CNNA.2005.1543200}.

\bibitem[Krizhevsky and Hinton(2009)]{krizhevsky2009learning}
A.~Krizhevsky and G.~Hinton.
\newblock Learning multiple layers of features from tiny images.
\newblock Technical report, Citeseer, 2009.

\bibitem[Kusuma and Fish(2005)]{doi:10.1190/1.1822449}
T.~Kusuma and B.~C. Fish.
\newblock Toward more robust neural‐network first break and horizon pickers.
\newblock In \emph{SEG Technical Program Expanded Abstracts 1993}, pages
  238--241, 2005.
\newblock \doi{10.1190/1.1822449}.
\newblock URL \url{https://library.seg.org/doi/abs/10.1190/1.1822449}.

\bibitem[Leggett et~al.(2003)Leggett, Sandham, and Durrani]{Leggett2003}
M.~Leggett, W.~A. Sandham, and T.~S. Durrani.
\newblock \emph{Automated 3-D Horizon Tracking and Seismic Classification Using
  Artificial Neural Networks}, pages 31--44.
\newblock Springer Netherlands, Dordrecht, 2003.
\newblock ISBN 978-94-017-0271-3.
\newblock \doi{10.1007/978-94-017-0271-33}.
\newblock URL \url{https://doi.org/10.1007/978-94-017-0271-33}.

\bibitem[Li et~al.(2012)Li, Ma, and Du]{A6184283}
L.~Li, G.~Ma, and X.~Du.
\newblock New method of horizon recognition in seismic data.
\newblock \emph{IEEE Geoscience and Remote Sensing Letters}, 9\penalty0
  (6):\penalty0 1066--1068, Nov 2012.
\newblock ISSN 1545-598X.
\newblock \doi{10.1109/LGRS.2012.2190039}.

\bibitem[Liu et~al.(2005)Liu, Xue, and Li]{doi:10.1190/1.1889749}
X.~Liu, P.~Xue, and Y.~Li.
\newblock Neural network method for tracing seismic events.
\newblock In \emph{SEG Technical Program Expanded Abstracts 1989}, pages
  716--718, 2005.
\newblock \doi{10.1190/1.1889749}.
\newblock URL \url{https://library.seg.org/doi/abs/10.1190/1.1889749}.

\bibitem[Meldahl et~al.(2001)Meldahl, Heggland, Bril, and
  de~Groot]{doi:10.1190/1.1438976}
P.~Meldahl, R.~Heggland, B.~Bril, and P.~de~Groot.
\newblock Identifying faults and gas chimneys using multiattributes and neural
  networks.
\newblock \emph{The Leading Edge}, 20\penalty0 (5):\penalty0 474--482, 2001.
\newblock \doi{10.1190/1.1438976}.
\newblock URL \url{https://doi.org/10.1190/1.1438976}.

\bibitem[Meldahl et~al.(2005)Meldahl, Heggland, Bril, and
  de~Groot]{doi:10.1190/1.1821262}
P.~Meldahl, R.~Heggland, B.~Bril, and P.~de~Groot.
\newblock The chimney cube, an example of semi‐automated detection of seismic
  objects by directive attributes and neural networks: Part i; methodology.
\newblock In \emph{SEG Technical Program Expanded Abstracts 1999}, pages
  931--934, 2005.
\newblock \doi{10.1190/1.1821262}.
\newblock URL \url{https://library.seg.org/doi/abs/10.1190/1.1821262}.

\bibitem[Poulton(2002)]{doi:10.1190/1.1484539}
M.~M. Poulton.
\newblock Neural networks as an intelligence amplification tool: A review of
  applications.
\newblock \emph{GEOPHYSICS}, 67\penalty0 (3):\penalty0 979--993, 2002.
\newblock \doi{10.1190/1.1484539}.
\newblock URL \url{https://doi.org/10.1190/1.1484539}.

\bibitem[Ronneberger et~al.(2015)Ronneberger, Fischer, and
  Brox]{Ronneberger2015}
O.~Ronneberger, P.~Fischer, and T.~Brox.
\newblock U-net: Convolutional networks for biomedical image segmentation.
\newblock \emph{Medical Image Computing and Computer-Assisted Intervention –
  MICCAI 2015}, page 234–241, 2015.
\newblock ISSN 1611-3349.
\newblock \doi{10.1007/978-3-319-24574-428}.
\newblock URL \url{http://dx.doi.org/10.1007/978-3-319-24574-428}.

\bibitem[Shi et~al.(2018)Shi, Wu, and Fomel]{doi:10.1190/segam2018-2997304.1}
Y.~Shi, X.~Wu, and S.~Fomel.
\newblock Automatic salt-body classification using deep-convolutional neural
  network.
\newblock In \emph{SEG Technical Program Expanded Abstracts 2018}, pages
  1971--1975, 2018.
\newblock \doi{10.1190/segam2018-2997304.1}.
\newblock URL
  \url{https://library.seg.org/doi/abs/10.1190/segam2018-2997304.1}.

\bibitem[Tingdahl and De~Rooij(2005)]{doi:10.1111/j.1365-2478.2005.00489.x}
K.~M. Tingdahl and M.~De~Rooij.
\newblock Semi-automatic detection of faults in 3d seismic data.
\newblock \emph{Geophysical Prospecting}, 53\penalty0 (4):\penalty0 533--542,
  2005.
\newblock \doi{10.1111/j.1365-2478.2005.00489.x}.
\newblock URL
  \url{https://onlinelibrary.wiley.com/doi/abs/10.1111/j.1365-2478.2005.00489.x}.

\bibitem[Veezhinathan et~al.(1993)Veezhinathan, Kemp, and
  Threet]{Veezhinathan1993}
J.~Veezhinathan, F.~Kemp, and J.~Threet.
\newblock A hybrid of neural net and branch and bound techniques for seismic
  horizon tracking.
\newblock In \emph{Proceedings of the 1993 ACM/SIGAPP Symposium on Applied
  Computing: States of the Art and Practice}, SAC '93, pages 173--178, New
  York, NY, USA, 1993. ACM.
\newblock ISBN 0-89791-567-4.
\newblock \doi{10.1145/162754.162863}.
\newblock URL \url{http://doi.acm.org/10.1145/162754.162863}.

\bibitem[Waldeland et~al.(2018)Waldeland, Jensen, Gelius, and
  Solberg]{doi:10.1190/tle37070529.1}
A.~U. Waldeland, A.~C. Jensen, L.-J. Gelius, and A.~H.~S. Solberg.
\newblock Convolutional neural networks for automated seismic interpretation.
\newblock \emph{The Leading Edge}, 37\penalty0 (7):\penalty0 529--537, 2018.
\newblock \doi{10.1190/tle37070529.1}.
\newblock URL \url{https://doi.org/10.1190/tle37070529.1}.

\bibitem[Wu and Zhang(2018)]{wu2018deep}
H.~Wu and B.~Zhang.
\newblock A deep convolutional encoder-decoder neural network in assisting
  seismic horizon tracking.
\newblock \emph{arXiv preprint arXiv:1804.06814}, 2018.

\bibitem[Wu and Fomel(2018)]{doi:10.1190/geo2017-0830.1}
X.~Wu and S.~Fomel.
\newblock Least-squares horizons with local slopes and multigrid correlations.
\newblock \emph{GEOPHYSICS}, 83\penalty0 (4):\penalty0 IM29--IM40, 2018.
\newblock \doi{10.1190/geo2017-0830.1}.
\newblock URL \url{https://doi.org/10.1190/geo2017-0830.1}.

\bibitem[Zhao(2018)]{doi:10.1190/segam2018-2997085.1}
T.~Zhao.
\newblock Seismic facies classification using different deep convolutional
  neural networks.
\newblock In \emph{SEG Technical Program Expanded Abstracts 2018}, pages
  2046--2050, 2018.
\newblock \doi{10.1190/segam2018-2997085.1}.
\newblock URL
  \url{https://library.seg.org/doi/abs/10.1190/segam2018-2997085.1}.

\end{thebibliography}
\end{document}